# Single-atom trajectories in higher-order transverse modes of a high-finesse optical cavity

T. Puppe, P. Maunz, T. Fischer, P.W.H. Pinkse, and G. Rempe
*Max-Planck-Institut für Quantenoptik, Hans-Kopfermann-Str. 1, D-85748 Garching, Germany*
*Email: gerhard.rempe@mpq.mpg.de*

Transits of single atoms through higher-order Hermite-Gaussian transverse modes of a high-finesse optical cavity are observed. Compared to the fundamental Gaussian mode, the use of higher-order modes increases the information on the atomic position. The experiment is a first experimental step towards the realisation of an atomic kaleidoscope.

Key words: cavity QED, cold atoms, optical cavities, transverse modes
PACS numbers: 42.50.Pq, 32.80.Pj, 42.60.Da

## 1. Introduction

A high-finesse optical cavity excited by a near-resonant laser beam is an ideal device to observe the motion of a single atom with high spatial and temporal resolution. It has been predicted, for example, that the position of a slowly moving atom can be measured with an accuracy well below the standard diffraction limit, defined as half the wavelength of the employed light [1, 2]. The measurement scheme is based on the fact that the presence of an atom in the cavity leads to a significant modification of both the amplitude and the phase of a weak light beam transmitted through the cavity. This modification is particularly large if the condition of strong coupling between the atom and the cavity is fulfilled, i.e. if both the atom and the cavity have a decay rate smaller than the characteristic frequency of the combined atom-cavity system, namely by the oscillatory exchange of a single quantum of energy between the atom and the cavity [3]. In this case, the transmitted intensity is a highly sensitive probe for the presence of an atom, so that efficient observation of a single moving atom becomes practical [4]. Indeed, three recent experiments performed with laser-cooled atoms have shown that it is possible to observe individual atoms with a time resolution of a few microseconds [5, 6, 7], an achievement hard to imagine for even the most sensitive optical microscope. The extraordinarily fast detection capability has subsequently been exploited to quickly increase the intensity of the probe light when the moving atom was detected at the cavity centre, thereby catching the observed atom in the cavity [8, 9]. A novel feature of these experiments is that the atom was trapped in a cavity field containing about one photon on average. In another experiment, a second independent laser beam was switched on to trap the atom in a far-detuned dipole trap of high intensity [10, 11, 12].

Without exception, all these experiments have utilised cavities operated in the fundamental Hermite-Gaussian $HG_{0,0}$ mode, which gives the strongest coupling of the atom to the cavity in the centre of an antinode. If the atom moves away from the centre, the coupling decreases. This results in a weaker influence on the cavity transmission. When averaged over the (fast) axial motion of the atom in the standing-wave field, the cavity transmission is therefore a good measure of the atom's radial position. However, because of the rotational symmetry of the $HG_{0,0}$ mode, angular position information cannot be obtained in a single measurement. Nevertheless, under certain constraints it was possible to estimate the atomic trajectory from the time evolution of the transmission by solving the classical equations of motion [9].

To determine the motion of an atom without these constraints, Horak et al. recently proposed a novel technique called the atomic kaleidoscope [13]. It takes advantage of frequency-degenerate higher-order transverse cavity modes to obtain two-dimensional position information from a single spatially resolved recording of the cavity transmission. The underlying idea is that an atom positioned off axis breaks the otherwise perfect rotational symmetry of the cavity. Two cases can be considered: First, an excited atom will emit into a unique superposition of transverse modes [14, 15]. This superposition called the effective mode depends on the position of the atom and maximises the atom-cavity coupling at the position of the atom. The spatial distribution of the light emitted from the cavity reflects the effective mode and therefore contains information about the atomic position in the plane perpendicular to the cavity axis. Second, if the empty cavity (without an atom) is excited by a laser beam, a light field with a certain spatial distribution builds up inside the cavity. This intra-cavity field is usually described as a superposition of different higher-order transverse $HG_{m,n}$ modes. But when an atom is present inside the cavity, it is more suitable to decompose the laser-excited mode into the effective mode and the residual mode which is orthogonal to the effective mode. By definition, the atom couples maximally to the effective mode. The residual mode is therefore not coupled to the atom. Hence, only the effective mode is affected by the presence of the atom, while the residual mode remains unaffected. This leads to a differential change of the amplitudes of the two modes from which the position of the atom can be determined by monitoring the spatial distribution of the light transmitted through the cavity.

Unfortunately, neither the method where the atom is excited nor the method where the cavity is excited has been demonstrated experimentally so far. The main reason is that measurements with the atomic kaleidoscope require a perfect rotational symmetry of the empty cavity in order to have a set of frequency-degenerate transverse modes. This imposes severe requirements on the quality of the mirrors.

In this paper, we therefore report on a new observation technique where two-dimensional information about the position of a moving atom is obtained without using frequency-degenerate modes. In fact, the frequency splitting of neighbouring higher-order modes induced by non-perfect mirrors is employed to subsequently excite these modes one-at-a-time by rapidly switching the frequency of a tuneable laser to the corresponding mode frequencies. The switching is so fast that the atom does not move an appreciable distance when monitoring the transmission of each mode. From the quasi-simultaneous transmission measurements performed on all these modes, the atomic trajectory in a plane perpendicular to the cavity axis can be determined.

We start by discussing briefly the parameters of our high-finesse cavity in chapter 2. Experimental data on single-atom transits through higher-order modes are presented in chapter 3. Finally, chapter 4 gives a conclusion and an outlook.

## 2. Cavity modes

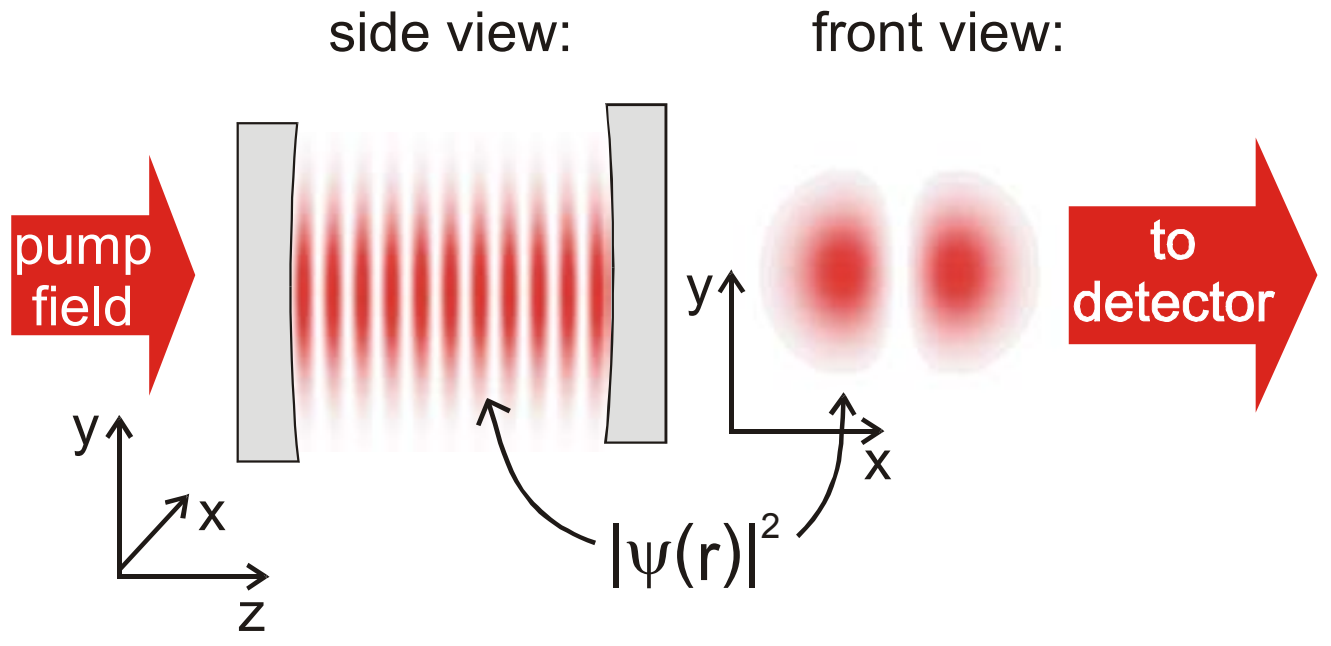

*Figure 1: Sketch of the optical standing-wave cavity with a measured $HG_{1,0}$ transverse mode depicted.*

The operational principle of the atomic kaleidoscope is based on a cavity with ideal spherical mirrors. Let us assume a general Fabry-Perot cavity as depicted in figure 1. Such a cavity supports many eigenmodes, which in the paraxial approximation can be described in a Hermite-Gaussian or Laguerre-Gaussian basis [16]. As the atom breaks the rotational symmetry, the Hermite-Gaussian basis, $HG_{m,n}$, is the natural choice. The standing wave along the cavity z-axis will give rise to a $\cos(2\pi z/\lambda)$ dependence of the mode, where $\lambda$ denotes the light wavelength. Since the axial motion is fast, it can be accounted for by averaging over the atom-cavity coupling in a suitable manner. The Hermite-Gaussian transverse mode functions can be described by

(1) $$\psi_{m,n}(x,y) = C_{m,n}\exp\left(\frac{-x^2-y^2}{w_0^2}\right)H_m\left(\frac{x\sqrt{2}}{w_0}\right)H_n\left(\frac{y\sqrt{2}}{w_0}\right)$$

where $w_0$ is the waist of the mode, m and n are the mode indices with respect to the x- and y-axes, $H_{m,n}$ are the corresponding Hermite polynomials of order m and n, and $C_{m,n}=(2^m\, 2^n\, m!\, n!)^{-1/2}\,(w_0^2\, \pi/2)^{-1/2}$ is a constant preserving the normalisation

(2) $$\int_{-\infty}^{\infty}\int_{-\infty}^{\infty}\psi^*_{m,n}(x,y)\psi_{m,n}(x,y)dxdy = 1 \ .$$

This normalisation indicates conservation of energy flow through a transverse plane in the cavity. With this normalisation, the coupling constant of an atom to the light field is $g_{m,n}(x,y)=g_0\ \psi_{m,n}(x,y)/\psi_{0,0}(0,0)$, where $g_0$ is the maximum coupling in the $HG_{0,0}$ mode. The intensity distribution of the modes is proportional to the square modulus of the mode function, $I_{m,n}(x,y)\sim|\psi_{m,n}(x,y)|^2$. The eigenfrequencies of these modes are

(3) $$\omega_{m,n,q} = \left[q + \frac{m+n+1}{\pi}\arccos\left(\sqrt{\xi_1\xi_2}\right)\right]\frac{\pi c}{L} \ ,$$

where q is the mode index along the cavity z-axis, $\xi_{1,2}=1-L/R_{1,2}$ with L the length of the cavity and $R_{1,2}$ the radius of curvature of the input and output mirror, respectively. It follows from equation (3) that modes of the same family, i.e. those with identical indices q and N=m+n, have the same eigenfrequency. Moreover, as the dielectric mirrors are assumed to show no polarisation effects, the electric field can be written as a scalar parameter. In other words, transverse modes with different polarisation are also frequency degenerate.

Real mirrors used in the laboratory, however, are never ideal. For mirrors with ultra-low losses forming a high-finesse cavity the acceptable tolerances are very small, as a small birefringence in the optical coating, for example, is amplified by the large number of reflections the light experiences inside the cavity. Small deviations from the perfect spherical shape result in frequency splittings of transverse mode families which are significant compared to the linewidth. We therefore briefly discuss the properties of our cavity relevant for the experiment described below. Further details in particular on isolating the mirrors from external vibrations can be found in [17].

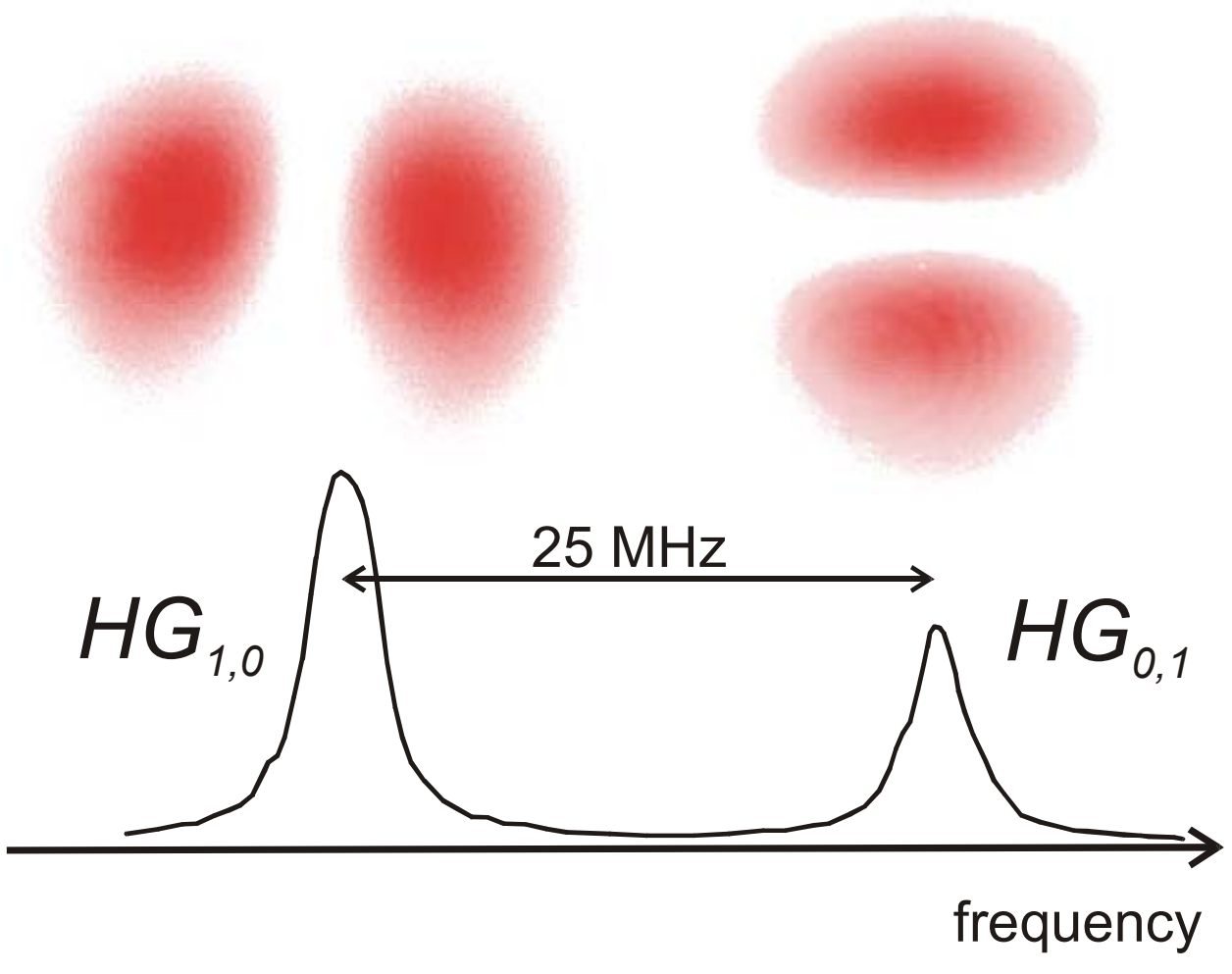

*Figure 2: Higher-order modes. The measured transverse intensity distributions of the $HG_{1,0}$ and $HG_{0,1}$ modes is shown in the upper part. For ideal spherical mirrors, the frequencies of these modes would be degenerate. The intensity transmitted through the cavity when scanning the frequency of the excitation laser is displayed in the lower part. The experimentally observed splitting is 25 MHz.*

The two mirrors forming the cavity are nominally identical. Both have a diameter of 7.75 mm and a radius of curvature of 20 cm. Their distance determines the length, L=123 µm, the free-spectral range, c/2L=1.22 THz, and the mode waist, $w_0$=29 µm, of our standing-wave cavity operated at the wavelength λ=780.2 nm. The finesse of the cavity is F=440,000, corresponding to a cavity linewidth of $2\kappa/2\pi$=2.8 MHz (FWHM).

The upper part of figure 2 shows the measured intensity pattern at the output mirror when a laser exciting the cavity is sequentially tuned to the frequencies of the two transverse modes of order N=1. The intensity distributions closely resemble the $HG_{1,0}$ and $HG_{0,1}$ modes, respectively. The slightly distorted form of the modes is probably due to aberrations of the imaging system. The modes are characterised by two bright regions separated by a dark nodal line. The nodal lines are oriented nearly horizontally or nearly vertically, respectively, neglecting a small tilt of about 5°. The orientation of the modes could be the result of a small deformation of the mirror surfaces from the ideal sphere, in first order described by two main axes that coincide with the nodal lines. It is accompanied by a mode splitting of 25 MHz, significantly larger than the linewidth, and independent of the polarisation of the input field. The lifting of the mode degeneracy could be explained by assuming two slightly different radii of curvature of the mirrors in the horizontal and vertical direction. It is worthwhile to mention that the splitting had changed to about 6 MHz when measured again some time later, although the cavity was never touched and never left the vacuum chamber during that period.

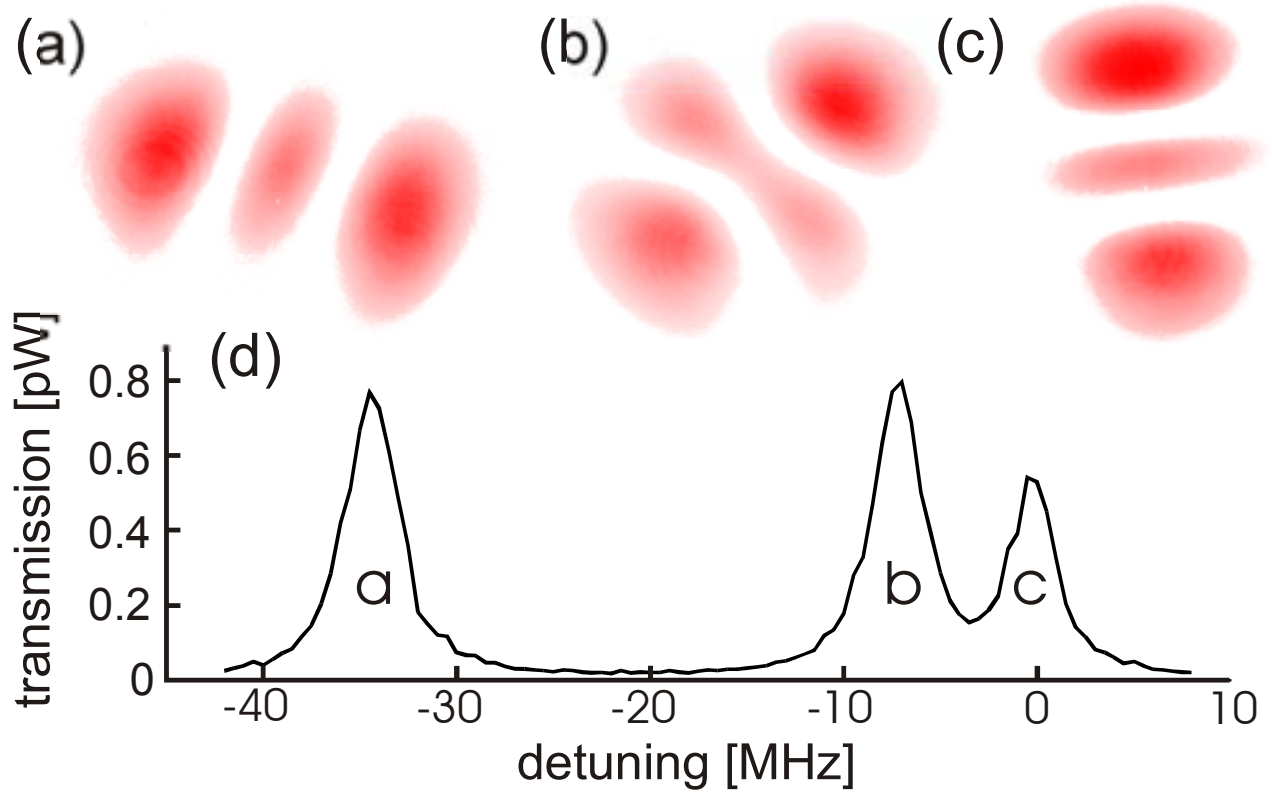

*Figure 3: Higher-order modes. The spatial pattern of the three modes of order N=2 are shown in (a)-(c). The resonance frequencies of the modes as determined by scanning the frequency of the excitation laser are indicated in (d).*

Note that a superposition of the two modes with N=1 is described by two parameters, which can be interpreted as (1) the orientation of the symmetry axes and (2) the deviation from rotational symmetry. The latter parameter describes a continuous transition between Laguerre-Gaussian and Hermite-Gaussian modes. With increasing mode index N>1, the number of parameters and, hence, the number of possible superpositions increases [14]. Therefore it is no surprise that these modes deviate noticeably from the standard Hermite-Gaussian basis modes. For instance, the three modes observed for N=2 are plotted in figure 3(a)-(c). These modes are characterised by three bright regions separated by two nodal lines. The intensity distributions of the three modes are similar, but rotated with respect to each other by an amount of about 60°. The frequencies of the three modes are distributed over an interval of about 34 MHz, as indicated in Fig. 3(d). The frequency of mode (a) is clearly separated from the frequencies of modes (b) and (c), of which the Lorentzian tails overlap significantly. A closer inspection shows that only mode (a) can be described to a very good approximation as a (rotated) $HG_{2,0}$ mode. Modes (b) and (c), in contrast, have a central bright region that differs from that of a $HG_{2,0}$ mode. This is particularly evident for mode (b) which tends to have some similarity with a (rotated) $HG_{1,1}$ mode due to the two intensity maxima located at the two ends of the central bright region.

For even higher transverse modes with N>2, the intensity distribution becomes very complex. It is questionable if these patterns can be explained by invoking just two different radii of curvature of the cavity mirrors.

## 3. Atomic transits

We now use the frequency non-degenerate modes of order N=1 to observe single atoms passing through the cavity in free flight. The technique of injecting laser-cooled atoms into the cavity has been described in detail elsewhere [6]. In short, a cloud of $^{85}$Rb atoms is collected from a background vapour in a magneto-optical trap and launched towards the cavity by means of a moving optical molasses. In this fountain geometry, the initial velocity of the atoms is determined by the final frequency difference of the laser beams forming the moving molasses. After being accelerated by the moving molasses the motion of the atoms is entirely determined by gravitational deceleration. Atoms with a near-zero velocity can be produced by tuning the classical turning point of the trajectory to the cavity centre. The horizontal velocity is always small, because of the geometrical selection by the narrow gap between the two mirrors and a differential pumping tube separating the chamber with the background vapour from the science chamber with the cavity.

We emphasise that, in principle, the force of the cavity-light field on a slowly moving atom can be quite significant, as has been studied in detail both experimentally [18] and theoretically [19]. In the experiments reported here, however, we use atoms with a large vertical velocity that dominates the horizontal velocity by far. Moreover, in most of the experiments we use light that is on resonance with the atomic transition. Hence, the (conservative) dipole force is zero. A numerical simulation of the experiment shows, however, that random momentum kicks caused by photon scattering leads to a significant increase of the atomic velocity in the direction of the cavity axis. To interpret the experimental data, theoretically calculated transmission signals must therefore be averaged over all possible atomic positions along the standing wave.

Before entering the cavity, the atoms are optically pumped into the $5S_{1/2}$ F=3, $m_F$=3 ground state. The probe laser exciting the cavity mode is circularly polarised and near-resonant with the atomic transition to the $5P_{3/2}$ F=4, $m_F$=4 state at a wavelength of 780.2 nm. In this way, an atomic two-level system is prepared. A small magnetic bias field with a magnitude below 1 Gauss and pointing along the cavity axis is applied to preserve the direction of the atom's magnetic moment in the vicinity of the cavity.

For the $HG_{0,0}$ mode and an atom at an anti-node on the axis of the cavity, the single-photon atom-cavity coupling constant amounts to $g_0/2\pi$=16 MHz, leading to a (vacuum-) Rabi frequency of 32 MHz. With a cavity-field decay rate of $\kappa/2\pi$=1.4 MHz and an atomic-dipole decay rate of $\gamma/2\pi$=3.0

MHz, the experiment is performed in the regime of strong coupling, defined by the condition g>(γ, κ). This condition is not violated even for higher-order transverse modes, as long as the size of the modes is not too large. For example, $g/2\pi=14$ MHz at both maxima of the $HG_{1,0}$ or $HG_{0,1}$ mode.

A well-stabilised laser is used to resonantly excite one or the other of the two $HG_{1,0}$ or $HG_{0,1}$ modes, as depicted in figure 2. The geometry of the pump field is adjusted to pump the modes with approximately equal strength. As discussed above, the frequency splitting of the two modes is so large that if the laser is tuned to the frequency of, e.g., the $HG_{1,0}$ mode, the excitation of the other $HG_{0,1}$ mode is suppressed by a factor of about 300. The excitation of the other mode can therefore safely be neglected. Launching atoms from below through the $HG_{1,0}$ mode with the nodal line oriented vertically then gives rise to a coupling variation with a single maximum. Consequently, the transit of the atom leads to a dip in the cavity transmission that is very similar to that of a transit through a $HG_{0,0}$ mode, as indicated in the upper part of figure 4. Choosing, however, a probe frequency resonant with the $HG_{0,1}$ mode gives rise to a double-peaked variation of the coupling strength. This is reflected in the double-dip structure of the transmission event shown in the lower part of figure 4. The increase of the cavity transmission at time 147.06 ms after the launch arises from the vanishing coupling when the atom crosses the horizontally oriented nodal line of the $HG_{0,1}$ mode.

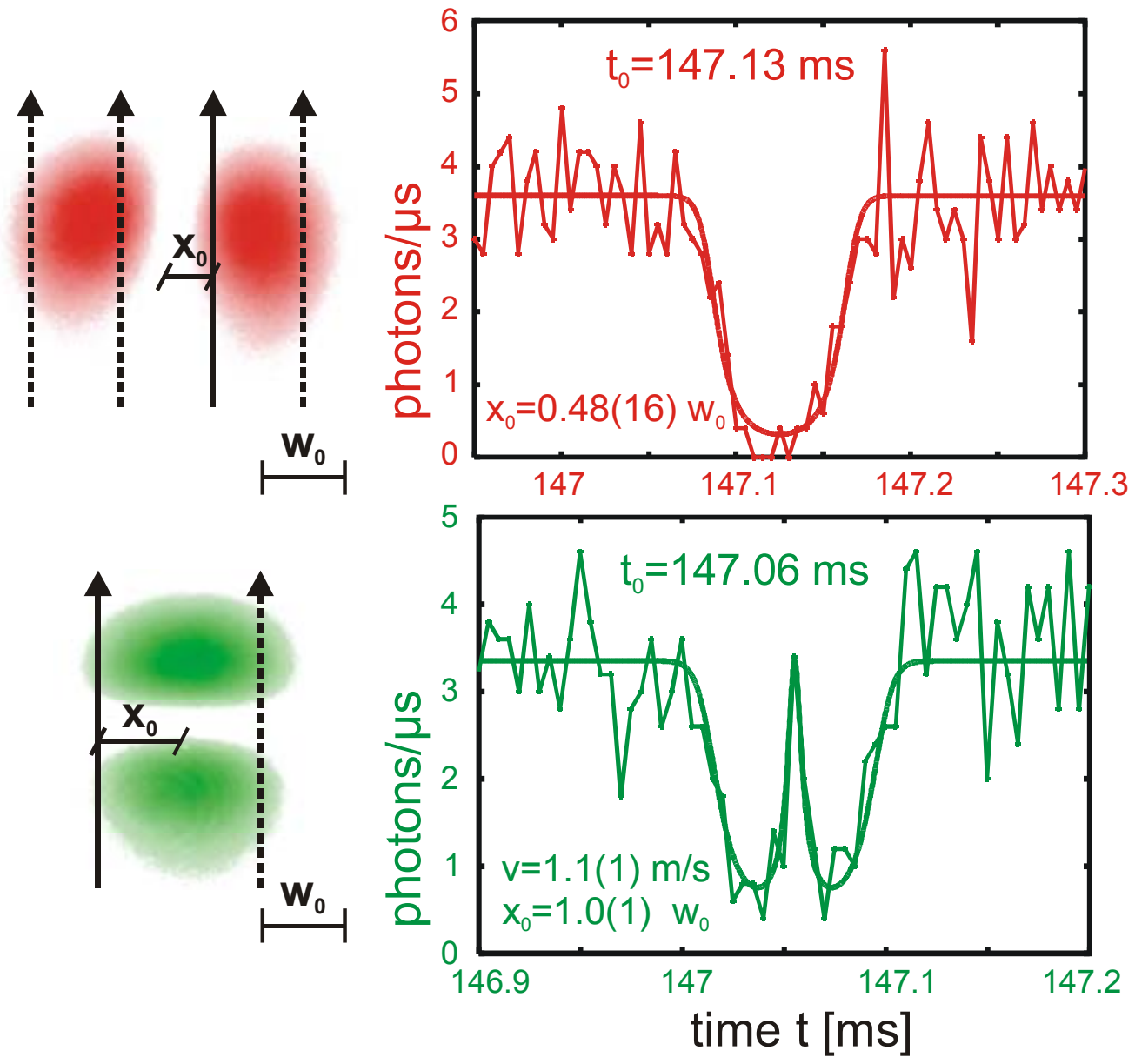

*Figure 4: Transit events. A single atom passing through a $HG_{1,0}$ mode (above) and a $HG_{0,1}$ mode (below) is observed by monitoring the transmission of the cavity. The probe laser is weak and resonant with both the depicted cavity mode and the atom, $\Delta_A=\Delta_C=0$. From a fit based on equation (4) (smooth curve), the arrival time of the atom at the centre of the cavity, $t_0$, the minimum distance to the cavity axis, $x_0$, and in case of the $HG_{0,1}$ mode also the vertical velocity, v, can be determined. The signal-to-noise ratio of the observed events is limited by shot noise caused by the small number of photons detected.*

More information about the atomic trajectory can be obtained by comparing the experimentally observed events with theoretically calculated signals. In the limit of a weak probe field leading to only a small excitation of the atom, as is the case in the experiment, the position-dependent transmission is given by

$$\frac{T(\vec{r})}{T_0}=\frac{\kappa^2\left(\Delta_A^2+\gamma^2\right)}{\left(\Delta_C\gamma+\Delta_A\kappa\right)^2+\left(g^2(\vec{r})-\Delta_A\Delta_C+\gamma\kappa\right)^2}, \tag{4}$$

where $T_0$ is the transmission of the empty cavity, and $\Delta_A=\omega_L-\omega_A$ and $\Delta_C=\omega_L-\omega_C$ denote the frequency difference between the laser frequency, $\omega_L$, and the atomic frequency, $\omega_A$, and the cavity frequency, $\omega_C$, respectively. As explained above, the transmission calculated from equation (4) must be averaged over the standing wave to take into account the fast light-induced motion of the atom along the cavity axis. The motion of the atom transverse to the cavity axis occurs with a vertical velocity v which is assumed to be not affected by the intra-cavity light field. The atom therefore follows a straight trajectory with a minimum distance to the cavity axis given by $x_0$.

For the $HG_{1,0}$ mode, $x_0$ and v cannot be considered independent, because both determine the duration of the transmission dip. We therefore fix $v\equiv v_b=1.07$ m/s as given by the arrival time and the ballistic trajectory of the atom after launching. The fitted point of closest approach to the cavity axis at t=147.13 ms is then $x_0=0.48(16)\ w_0$. But note that obviously there are three more possible solutions (given by $x_0=-0.48\ w_0$ and $x_0=\pm1.64\ w_0$, respectively) that would yield an equivalent transit signal.

In contrast to the $HG_{1,0}$ mode, the vertical distance between the two intensity maxima of the $HG_{0,1}$ mode, $2^{1/2}\ w_0$, forms a natural ruler. In this case it is possible to independently determine the vertical velocity and the point of closest approach, with the results v=1.1(1) m/s and $x_0=1.0(1)$ $w_0$ at t=147.06 ms. The velocity v agrees with the one expected from the ballistic trajectory, $v_b=1.07$ m/s. Note that there is only one other trajectory (with $x_0=-1.0\ w_0$) which would lead to an equivalent transmission signal.

We emphasise that the two transit signals displayed in figure 4 for the two modes are produced by two different atoms from two different fountain launches. A much better way to track the atomic motion is to use both modes simultaneously. Given a sufficiently high signal-to-noise ratio, this would allow one to determine the position of the atom from a single measurement at one time only. Note, however, that the transmitted intensity on any given mode is only a measure for the magnitude of the coupling constant, g. Therefore any measurement of the atomic position in a plane perpendicular to the cavity axis is restricted to contour lines of equal coupling constant. For the case of the two $HG_{1,0}$ and $HG_{0,1}$ modes with N=1, this leads to in general 8 equivalent positions defined by the interception of the 4 possible contour rings. This degeneracy can be reduced by using a larger number of modes, i.e. modes of a family with N>1. Moreover, the position resolution increases with N. But note, that two point-symmetric positions cannot be distinguished unless modes of odd and even N are used. A suitable (quasi-) degeneracy of such modes can be achieved by tuning the length of a near-planar cavity such that the frequencies of mode families with different longitudinal mode order q coincide.

The measurement scheme just outlined could be performed by simultaneous excitation of all relevant modes and observation of their transmissions by spatial filtering of the cavity output with, e.g., an ultra-sensitive camera. In the case of non-degenerated modes, frequency filtering could be implemented by heterodyne detection. A simplified version of the latter can be realised by quickly alternating the laser frequency between the modes, thereby exciting only one mode at any given time. In that way, the cavity transmission can be identified with the corresponding mode from the known timing of the frequency of the probe laser.

To demonstrate the method, we again use the $HG_{1,0}$ and the $HG_{0,1}$ modes. For rapid switching, the laser light is passed through an acousto-optic modulator connected to two different radio frequency sources by means of a fast switch which alternatively turns on one of the sources while turning off the other one. Switching back and fro between the two modes is performed at a frequency of 200 kHz. This gives a time interval of 2.5 µs for probing the transmission of each mode. This time interval is short compared to the approximately 100 µs long transit time of the atom through the cavity. Note that the light power inside the cavity needs about 300 ns to equilibrate after switching. All photons detected during this time interval after switching are discarded in order not to distort the signal. In this way, two-dimensional position information is obtained from the quasi-simultaneous transmission measurement of two cavity modes.

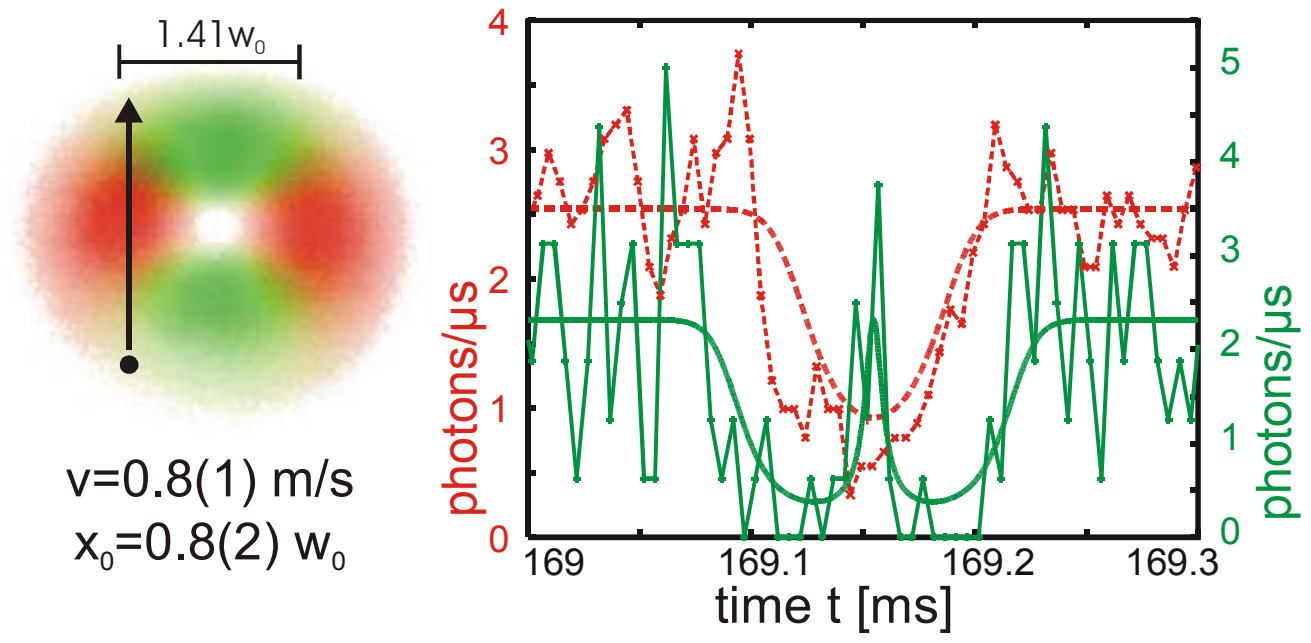

*Figure 5: Transit of a single atom through alternatively excited $HG_{1,0}$ and $HG_{0,1}$ modes. The detunings are $\Delta_A/2\pi$=-25 MHz, $\Delta_C$=0 for the $HG_{1,0}$ mode, and $\Delta_A$=$\Delta_C$=0 for the $HG_{0,1}$ mode. The left (right) vertical axis denotes the photon flux at the $HG_{1,0}$ ($HG_{0,1}$) mode frequency. The noise is shot noise due to the small photon flux limited by the condition of weak probing. Fitting a constant-velocity vertical trajectory to the experimental data, the velocity, v=0.8(1) m/s, and the minimum distance to the cavity axis, $x_0$=0.8(2)$w_0$, and, is obtained.*

The experimental result for an atom crossing the cavity close to its axis is depicted in figure 5. Clearly, the number of photons detected is too small, and hence the signal too noisy, to determine the position of the atom from a single transmission measurement with both the $HG_{1,0}$ and the $HG_{0,1}$ modes. However, for the displayed transit event, enough information is available to estimate the velocity and the horizontal position of the trajectory. Analogous to the single-mode case, a fit to the transmission signal of the $HG_{0,1}$ mode determines the velocity, v=0.8(1) m/s, the point of closest approach, $x_0$=0.8(2) $w_0$, and the arrival time t=169.15 ms. The velocity agrees well with the expected $v_b$=0.72 m/s, and the deduced transmission signal for the other $HG_{1,0}$ mode calculated without any further free parameter matches the experimental data reasonably well. But note that the experimentally observed transmission of the $HG_{1,0}$ mode tends to run below the theoretical expectation. This could be explained by invoking a higher atom-cavity coupling, consistent with the fact that in the experiment the laser and the cavity are both red detuned from the atomic transition by an amount of 25 MHz, with the result that the dipole force attracts the atom towards the antinodes of the cavity.

Note that the two transit signals displayed in figure 5 are more noisy than the corresponding signals in figure 4. This is a simple consequence of the fact that, for the same time resolution, the switching leads to a more than twofold reduction of the measurement time, which could be considered the cost for ease of operation when compared with, e.g., a heterodyne technique.

We now briefly address the question of using modes of very high order to observe the motion of an atom. A problem here is that the larger mode volume decreases the maximal atom-cavity coupling constant. Hence, the contrast of the transit signals through standard $HG_{m,n}$ modes will decrease. Moreover, as the mode patterns become more and more complex, it will be increasingly difficult to find modes with well-resolved resonance frequencies, a necessity for the interpretation of the transmission signals. By coincidence, at N=10 a stretched mode reminiscent of the $HG_{0,10}$ mode with nodal lines oriented horizontally was found in our cavity. Unfortunately, it turned out that single transits do not have enough contrast to provide meaningful data. We therefore calculate from the stream of photons detected during many single-atom transits the intensity autocorrelation functions, $g^{(2)}(\tau)$. The autocorrelation function is defined as

$$g^{(2)}(\tau)=\left\langle \frac{P(t,t+\tau)}{P(t)P(t+\tau)} \right\rangle , \quad (5)$$

where P(t, t+τ) is the probability to detect a photon at time t+τ provided a photon was observed at time t. The unconditioned probability for a photon-detection event at time t is P(t). The entire function is averaged over all available times t. Note that a characteristic frequency *f* in the transit signal of a single atom shows up as an oscillation in $g^{(2)}(\tau)$ with a characteristic temporal spacing of 1/*f*. Hence, the autocorrelation function of different transits displaying the same characteristic frequency can be averaged, thereby increasing the signal-to-noise ratio. This provides a simple method to visualise characteristic oscillations in the transmission of the cavity that are not visible in single transits [18].

An experimental result is shown in figure 6. The oscillations displayed in the picture result from the spatial structure of the $HG_{0,10}$ mode. As expected, many local maxima are visible in the autocorrelation function, which settles to a constant value at times exceeding 120 µs, the transit time of the atoms through the mode. The non-perfect orientation and the shape of the mode can be an explanation for the fact that not all of the expected 11 local maxima are clearly visible. An alternative explanation might be that the atoms are deflected from their straight, constant-velocity trajectory by light forces.

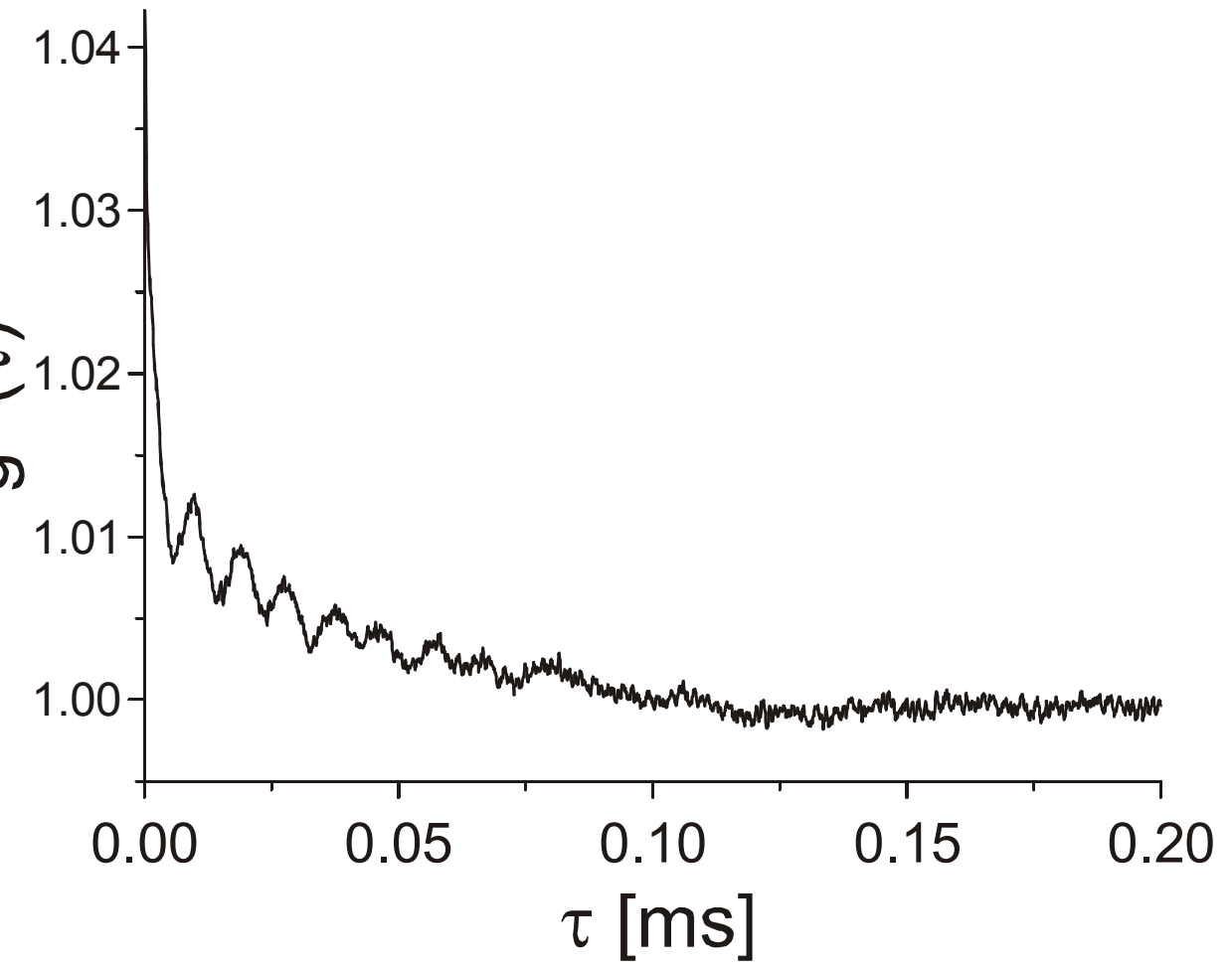

*Figure 6: Autocorrelation function of single-atom transits through a $HG_{0,10}$ mode. The detuning of the laser from the atom and the cavity are $\Delta_A/2\pi$=-50 MHz and $\Delta_C/2\pi$=-1.5 MHz, respectively. The motion of the atoms through many of the 11 local intensity maxima of the mode can clearly be discriminated. The function is normalised to approach 1 for large $\tau$.*

In fact, light forces are another motivation for looking at transverse modes with high N, as they could enhance cavity-mediated cooling forces along the radial direction, perpendicular to the cavity axis. Consider, for example, a $HG_{N,0}$ mode for which the maximum coupling in each of the N»1 intensity maxima scales as $N^{-1/4}$. As the size of the mode scales as $N^{1/2}$, the distance between neighbouring mode maxima scales as $N^{-1/2}$. Therefore the gradient of the coupling constant increases with N as $N^{1/4}$. As cooling forces are proportional to the gradient of the coupling constant squared, it is favourable to use higher order transverse modes. Indeed, Salzburger et al. [20] have proposed to use the light forces in higher-order transverse cavity modes to enhance the capturing of atoms in a high-finesse cavity.

## 4. Conclusion and outlook

Higher-order transverse cavity modes are an ideal tool to obtain in a direct manner information about the motion of a single atom in a plane perpendicular to the axis of a high-finesse optical cavity. The technique is superior to previous methods based on the standard $HG_{0,0}$ mode. Our technique is illustrated by a first experiment that exploits the frequency splitting of two low-order transverse modes to monitor the motion of a freely moving atom. In another experiment, the motion of a single atom through a $HG_{10,0}$ mode was observed, demonstrating that atoms can be made to interact with high-order transverse modes.

The position information obtained from the atomic motion can be improved by using higher-order transverse modes in a smaller cavity with a larger atom-cavity coupling constant. Moreover, selecting specific mirrors and changing their relative orientation should allow one to tailor the mode spectrum in such a way that a specific family of high-order modes can be used to observe the atomic motion. It should even be possible to remove the splitting and implement the atomic kaleidoscope with a cavity supporting frequency-degenerate modes.

Higher-order transverse modes might also be used to enhance cavity-mediated light forces. Alternatively, it might be more effective to measure the transverse position of an atom trapped in a far-detuned intra-cavity light field and apply feedback on the atomic motion for cooling. A first experiment in this direction has already been performed recently [21]. A dipole trap formed by a symmetry-breaking higher-order mode with an amplitude controlled by fast feedback circuitry should allow one to cool both the radial and the angular motion of the orbiting atom.

---

**References**

[1] Rempe, G., Appl. Phys. B **60** 233 (1995).
[2] Quadt, R., Collett, M., Walls, D.F., Phys. Rev. Lett. **74** 351 (1995).
[3] For a review, see, e.g. “Cavity Quantum Electrodynamics”, Advances in Atomic, Molecular, and Optical Physics **2**, edited by P.R. Berman, (Academic Press, 1994).
[4] Pinkse, P.W.H., Fischer, T., Maunz, P., Puppe, T., and Rempe, G., J. Mod. Optics **47** 2769 (2000).
[5] Mabuchi, H., Turchette, Q.A., Chapman, M.S., and Kimble, H.J., Opt. Lett. **21** 1393 (1996).
[6] Münstermann, P., Fischer, T., Pinkse, P.W.H., and Rempe, G., Opt. Commun. **159** 63 (1999).
[7] Sauer, J.A., Fortier, K.M., Chang, M.S., Hamley, C.D., and Chapman, M.S., quant-ph/0309052.
[8] Pinkse, P.W.H., Fischer, T., Maunz, P., and Rempe, G., Nature **40** 365 (2000).
[9] Hood, C.J., Lynn, T.W., Doherty, A.C., Parkins, A.S., and Kimble, H.J., Science **28** 1447 (2000).
[10] Ye, Y., Vernooy, D.W., and Kimble, H.J., Phys. Rev. Lett. **83** 4987 (1999).
[11] McKeever, J., Buck, J.R., Boozer, A.D., Kuzmich, A., Nägerl, H.-C., Stamper-Kurn, D.M., and Kimble, H.J., Phys. Rev. Lett. **90** 133602 (2003).
[12] Maunz, P. et al. (to be published).
[13] Horak, P., Ritsch, H., Fischer, T., Maunz, P., Puppe, T., Pinkse, P.W.H., Rempe, G., Phys. Rev. Lett. **88** 043601 (2002).
[14] Maunz, P., Puppe, T., Fischer, T., Pinkse, P.W.H., and Rempe, G., Opt. Lett. **28** 46 (2003).
[15] T. Fischer, PhD thesis, Technische Universität München (2002), http://tumb1.biblio.tu-muenchen.de/publ/diss/ph/2003/fischer.html.
[16] Siegman, A.E., Lasers (University Science Books, 1986).
[17] Pinkse, P.W.H. and Rempe, G., in "Cavity-Enhanced Spectroscopies", Experimental methods in the physical sciences, Vol. 40, edited by R.D. van Zee and J.P. Looney, Academic Press (2002).
[18] Münstermann, P., Fischer, T., Maunz, P., Pinkse, P.W.H., and Rempe, G., Phys. Rev. Lett. **82** 3791 (1999).
[19] Domokos, P., and Ritsch, H., J. Opt. Soc. Am. B **20** 1098 (2003).
[20] Salzburger, T., Domokos, P., and Ritsch, H., Opt. Expr. **10** 1204 (2002).
[21] Fischer, T., Maunz, P., Pinkse, P.W.H., Puppe, T., and Rempe, G., Phys. Rev. Lett. **88** 163002 (2002).